\documentclass[aps,prl,reprint,longbibliography,notitlepage,superscriptaddress,preprintnumbers]{revtex4-1}

\usepackage{graphics}      
\usepackage{graphicx}      
\usepackage{url}           
\usepackage{bm}            
\usepackage{amsmath}
\usepackage{amssymb}
\usepackage{physics}
\usepackage[caption=false]{subfig}

\usepackage{tikz}
\usetikzlibrary{snakes}

\usepackage[colorlinks=true, pdfstartview=FitV, linkcolor=red, citecolor=blue, urlcolor=blue]{hyperref}
\usepackage{xcolor}
\usepackage[utf8]{inputenc}

\newcommand{\Yobk}{Y_{0,\mathrm{BK}}}
\newcommand{\xbj}{x_\mathrm{Bj}}

\newcommand{\nc}{N_\mathrm{c}}
\newcommand{\cf}{C_\mathrm{F}}
\newcommand{\as}{\alpha_\mathrm{s}}
\newcommand{\gev}{\,\mathrm{GeV}}

\newcommand{\xt}{{\mathbf{x}}}

\newlength{\mycol}
\usepackage{capt-of}

\begin{document}

\title{Proton structure functions at NLO in the dipole picture with massive quarks}

\preprint{HIP-2022-28/TH}

\author{Henri Hänninen}
\email{henri.j.hanninen@jyu.fi}
\affiliation{ Department of Mathematics and Statistics, University of Jyv\"askyl\"a,
P.O. Box 35, 40014 University of Jyv\"asky\"a, Finland}
\affiliation{ Department of Physics, University of Jyv\"askyl\"a,
P.O. Box 35, 40014 University of Jyv\"asky\"a, Finland}
\affiliation{ Helsinki Institute of Physics, P.O. Box 64, 00014 University of Helsinki, Finland}

\author{Heikki Mäntysaari}

\email{heikki.mantysaari@jyu.fi  }
\affiliation{ Department of Physics, University of Jyv\"askyl\"a,
P.O. Box 35, 40014 University of Jyv\"asky\"a, Finland}
\affiliation{ Helsinki Institute of Physics, P.O. Box 64, 00014 University of Helsinki, Finland}

\author{Risto Paatelainen}
\email{risto.paatelainen@helsinki.fi}
\affiliation{ Helsinki Institute of Physics, P.O. Box 64, 00014 University of Helsinki, Finland}

\author{Jani Penttala}
\email{jani.j.penttala@jyu.fi}
\affiliation{ Department of Physics, University of Jyv\"askyl\"a,
P.O. Box 35, 40014 University of Jyv\"asky\"a, Finland}
\affiliation{ Helsinki Institute of Physics, P.O. Box 64, 00014 University of Helsinki, Finland}

\begin{abstract}
We predict heavy quark production cross sections in Deep Inelastic Scattering at high energy by applying the Color Glass Condensate effective theory. We demonstrate that when the calculation is performed consistently at next-to-leading order accuracy with massive quarks it becomes possible, for the first time in the dipole picture with perturbatively calculated center-of-mass energy evolution, to simultaneously describe both the light and heavy quark production data at small $\xbj$.
We furthermore show how the heavy quark cross section data provides additional strong constraints on the extracted non-perturbative initial condition for the small-$\xbj$ evolution equations.  
\end{abstract}

\maketitle

\emph{Introduction} --- %
Probing the properties of the non-linearly behaving gluonic matter in protons and nuclei at high energies is a major science goal of the future Electron-Ion Collider (EIC)~\cite{AbdulKhalek:2021gbh,Aschenauer:2017jsk,Accardi:2012qut}. 
Measuring the total and heavy quark production cross sections in Deep Inelastic Scattering (DIS) off nuclei is especially intriguing, as non-linear saturation effects are enhanced in heavy nuclei~\cite{Kowalski:2007rw}. The EIC will be able to perform very precise total cross section measurements over a relatively wide kinematical domain characterized by the gluon longitudinal momentum fraction $\xbj$ 
and the photon virtuality $Q^2$. 

Non-linear gluon saturation effects are expected to have  a modest effect on structure functions in the EIC kinematics (see e.g.~\cite{Marquet:2017bga,Mantysaari:2018nng,Armesto:2022mxy}). 
To unambiguously determine the existence of non-linear QCD dynamics at collider energies and to quantify its role on the small-$\xbj$ structure of protons and nuclei, it is likely necessary to perform a global analysis of the future proton and nuclear DIS data at small $\xbj$. In particular, it will be important to include both the inclusive and heavy quark production data that have different sensitivities on saturation effects in order to extract in detail the properties of the QCD matter at extremely large parton densities. Charm production is an especially powerful process as the charm mass  is large enough to suppress non-perturbative effects, but simultaneously light enough to allow one to access QCD dynamics in the non-linear regime.

To describe QCD dynamics at high energies, where parton densities are very large and emergent non-linear phenomena dominate, it is convenient to use the Color Glass Condensate (CGC)~\cite{Gelis:2010nm,Weigert:2005us} effective field theory framework. 
The DIS process is then naturally described in the dipole picture~\cite{Nikolaev:1990ja,Mueller:1994gb}, 
where 
the photon splits into a quark-antiquark pair long before the interaction with the target. The interaction of the quark dipole with the target is then taken to be eikonal, i.e. the transverse coordinates of the partons do not change when they traverse through the target color field. In this picture, leading-order (LO) calculations including a resummation of the high-energy logarithms $\as \ln 1/x$ to all orders (where $\as$ is the strong coupling) within the CGC framework have been successful in describing the precise proton structure function data from HERA~\cite{Albacete:2010sy,Lappi:2013zma,Ducloue:2019jmy}. This suggests that the HERA data is compatible with the hypothesis that gluon saturation is manifest at HERA energies. In addition, calculations based on collinear factorization 
have also found the resummation of the high-energy logarithms to be important in order to describe the details of the HERA data~\cite{Ball:2017otu}.

The structure function data is  used to constrain the non-perturbative initial condition for the small-$\xbj$ evolution equations. Therefore, a good description of the total cross section data is crucial when applying the CGC framework to describe any other scattering process (e.g. proton-nucleus collisions at the LHC~\cite{Tribedy:2011aa,Lappi:2013zma,Stasto:2013cha,Albacete:2016tjq,Mantysaari:2019nnt,Shi:2021hwx}).
Compatibility with the available cross section data is also required when
developing a realistic description for the early stages of heavy-ion collisions~\cite{Schenke:2012wb}, needed to extract the fundamental properties of the Quark-Gluon Plasma.

 In this Letter, we present predictions for  heavy quark production cross sections in DIS using the non-perturbative initial condition for the perturbative Balitsky-Kovchegov (BK) small-$\xbj$  evolution equation~\cite{Kovchegov:1999yj,Balitsky:1995ub}, 
determined from the fits to total DIS cross section data in~\cite{Beuf:2020dxl}. The predicted heavy quark cross sections are shown to be in excellent agreement with the HERA  data~\cite{H1:2018flt}. This is the first time in the CGC framework that a simultaneous description of total and heavy quark production data is achieved in calculations where the energy dependence is obtained by solving the small-$\xbj$ evolution equation. 
A crucial ingredient here is the
next-to-leading order (NLO) accuracy in $\as$ recently achieved for the massive impact factors from first-principle light-cone perturbation theory calculations~\cite{Beuf:2022ndu,Beuf:2021srj,Beuf:2021qqa}.   
We also demonstrate how the
heavy quark production data can provide additional
constraints for the extracted non-perturbative initial condition of the BK evolution.

The results presented here are from the first-ever numerical calculation of the heavy quark structure functions in the dipole picture at NLO.
The successful description of the HERA data demonstrates that future global analyses
are feasible 
and can be applied to probe in detail gluon saturation at the LHC and future EIC, where nuclear targets with larger saturation scales are available.

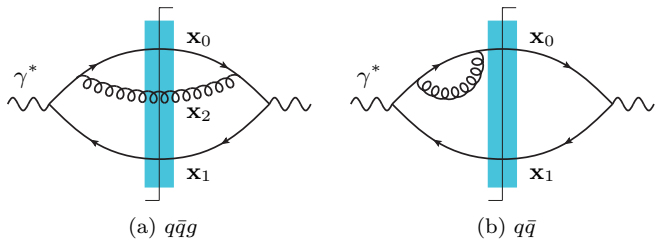
\begin{figure}
     \centering
    \subfloat[$q\bar q g$\label{fig:qqg}]{
         \centering
         \includegraphics[width=0.49\columnwidth]{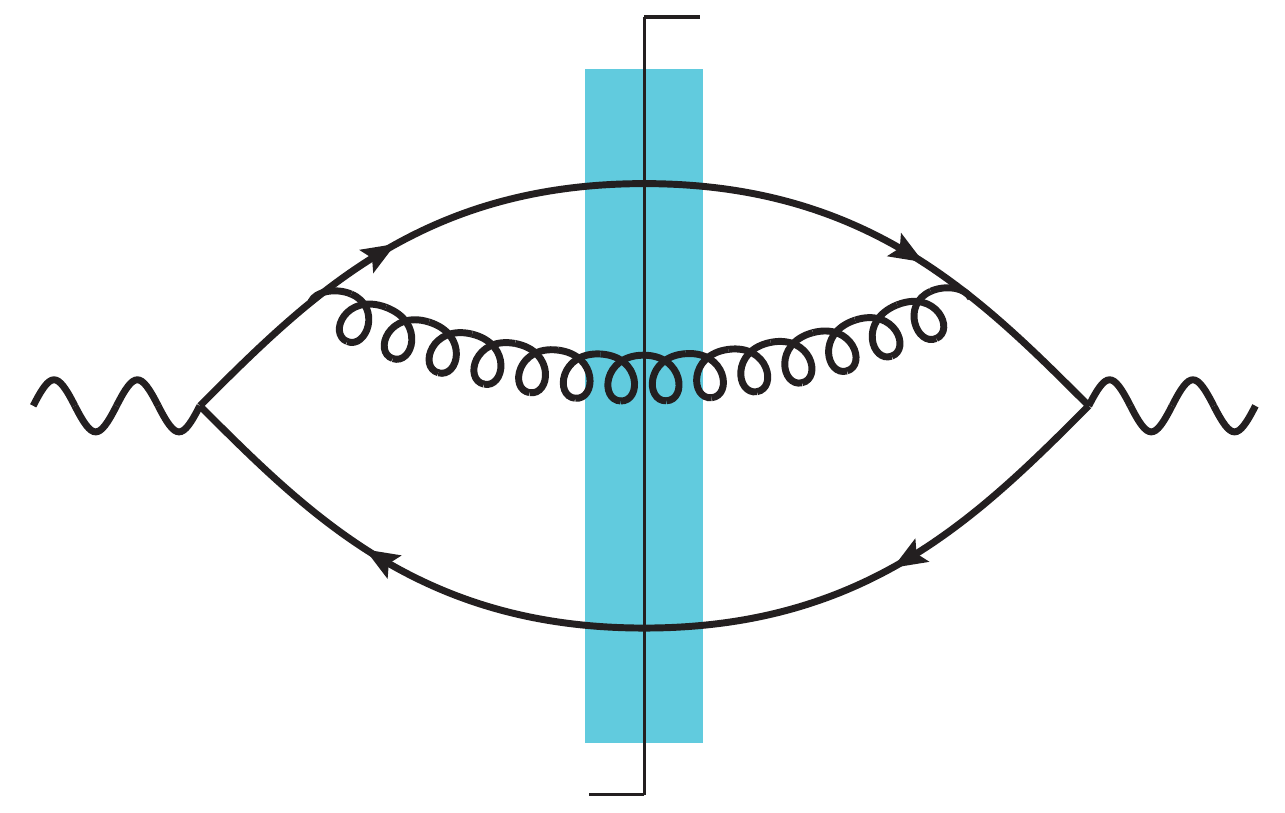}
          \begin{tikzpicture}[overlay]
         \node[anchor=south east] at (-1.4cm,2.cm) {$\xt_0$};
         \node[anchor=south east] at (-1.4cm,0.2cm) {$\xt_1$};
         \node[anchor=south east] at (-1.4cm,1.cm) {$\xt_2$};
        \node[anchor=south east] at (-3.7cm,1.45cm) {$\gamma^*$};
        \end{tikzpicture}
         }
    \subfloat[$q\bar q$\label{fig:qq}]{
         \centering
         \includegraphics[width=0.49\columnwidth]{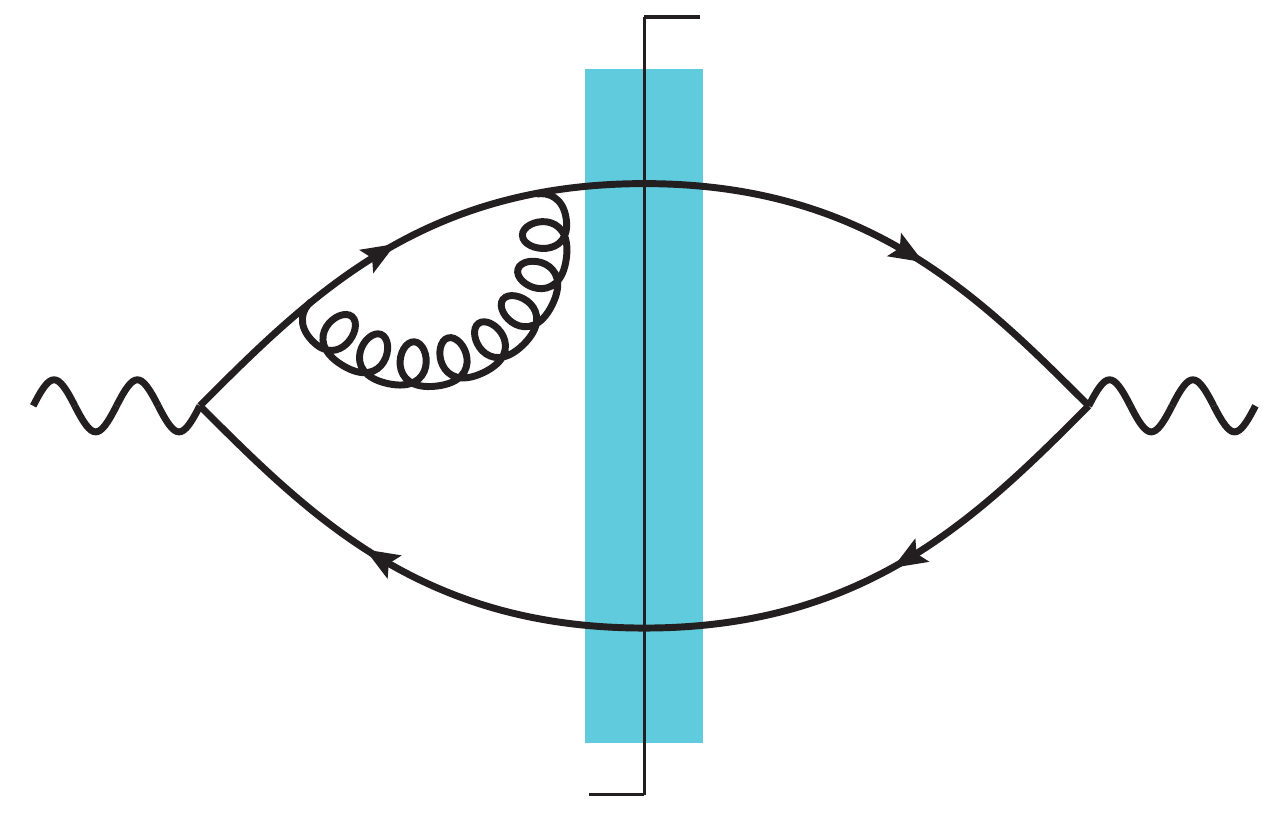}
         \begin{tikzpicture}[overlay]
         \node[anchor=south east] at (-1.4cm,2.0cm) {$\xt_0$};
         \node[anchor=south east] at (-1.4cm,0.2cm) {$\xt_1$};
         \node[anchor=south east] at (-3.7cm,1.45cm) {$\gamma^*$};
         \end{tikzpicture}
    }
     \caption{
    Example diagrams contributing to the elastic $\gamma^*p$ amplitude at NLO. The blue band represents the dipole--shockwave interaction. 
     }
\end{figure}

\emph{Structure functions at high energy} --- %
Using the optical theorem, the total virtual photon $(\gamma^*)$ -- proton ($p$) cross section can be obtained from the forward elastic $\gamma^* + p \to \gamma^* + p$ scattering amplitude. In the dipole picture, the $\gamma^* + p$ scattering 
is described in terms of eikonal interactions between the  partonic Fock states of the photon and the target color field, and perturbatively calculable \emph{impact factors} describing the photon fluctuations to the given partonic states. 
Eikonal interactions with the target are encoded in the Wilson lines, which are the scattering matrix elements for bare partons propagating through the target color field. 

At NLO  the contributing photon Fock states are the quark-antiquark $|q\bar q\rangle$ and quark-antiquark-gluon $|q\bar q g\rangle$ states. Therefore, at NLO the total virtual photon cross section can be schematically decomposed into two parts. The first contribution (illustrated in Fig.~\ref{fig:qqg}) corresponds to the case where the $q\bar q g$ system crosses the shockwave:
\begin{equation}
    \sigma^{\gamma^*}_{q\bar qg} = K_{q\bar qg} \otimes N_{012}. \label{eq:sigma_qqg} 
\end{equation}
The second contribution (illustrated in Fig.~\ref{fig:qq}), which includes the lowest-order part (interaction with an unevolved target) and the one-gluon-loop QCD corrections to it, reads 
\begin{equation}
\label{eq:sigma_qq}
    \sigma^{\gamma^*}_{q\bar q} = K_{q\bar q} \otimes N_{01}.
\end{equation}
Here $K_{q\bar q}$ and $K_{q\bar q g}$ refer to the perturbatively computed NLO impact factors
obtained with massive quarks in~\cite{Beuf:2022ndu,Beuf:2021srj,Beuf:2021qqa} and in the massless quark limit in~\cite{Hanninen:2017ddy,Beuf:2016wdz,Beuf:2017bpd}. In addition, the notation $\otimes$ refers to an integral over the parton transverse coordinates $\xt_i$ and longitudinal momentum fractions in the mixed space. 
Additionally, $N_{01}$ and $N_{012}$ are correlators of two or three Wilson lines,
where the subscripts $0,1,2$ refer to the transverse coordinates of the quark, antiquark and the gluon. In terms of the  Wilson lines $V(\xt)$ in the fundamental representation these correlators read:
\begin{align}
    S_{01} &= \frac{1}{\nc} \left\langle  \Tr{V(\xt_0) V^\dagger(\xt_1)} \right\rangle , \label{eq:s01}\\
    S_{012} &= \frac{\nc}{2\cf} \left( S_{02}S_{21} - \frac{1}{\nc^2} S_{01}\right). \label{eq:s012}
\end{align}
Here  $\langle \cdots \rangle$ refers to the average over the target color charge configurations, $\nc$ is the number of colors, $\cf = (\nc^2-1)/(2\nc)$, $S_{ij}=1-N_{ij}$  and $S_{ijk}=1-N_{ijk}$. In addition, we have used the mean-field limit (which is  a precise approximation~\cite{Kovchegov:2008mk}) to factorize the expectation value of the product to a product of expectation values.

The Wilson lines and their correlators satisfy small-$\xbj$ evolution equations describing their dependency on the center-of-mass energy (see~\cite{Ducloue:2019ezk} for a detailed discussion of the evolution variable).
The dipole amplitude $N_{01}$ satisfies the BK 
equation~\cite{Kovchegov:1999yj,Balitsky:1995ub} and via Eq.~\eqref{eq:s012}  $N_{012}$ also depends on the center-of-mass energy. The evolution rapidity depends on the lower limit of the emitted gluon longitudinal momentum fraction~\cite{Ducloue:2017ftk,Beuf:2020dxl}. 
The integration over the emitted gluon phase space in Eq.~\eqref{eq:sigma_qqg} contributes a large logarithm of energy that modifies the scattering amplitude of the original dipole $N_{01}$. These logarithms are resummed into the BK equation~\cite{Ducloue:2017ftk}.
The BK  equation and a numerical solution to it are  known at NLO~\cite{Balitsky:2008zza,Lappi:2015fma,Lappi:2016fmu}. 
We use the initial condition fitted to the HERA data in~\cite{Beuf:2020dxl}  including only massless quarks, where the full (numerically heavy) NLO BK equation has been approximated by evolution equations that use different schemes to resum the most important higher-order corrections. 
The same evolution equations, \emph{ResumBK}~\cite{Iancu:2015vea,Iancu:2015joa}, \emph{KCBK}~\cite{Beuf:2014uia} and \emph{TBK}~\cite{Ducloue:2019ezk} referring to different resummation schemes, are used in this work as in~\cite{Beuf:2020dxl}.

The structure functions are written in terms of the total virtual photon-target cross sections as $    F_2 = \frac{Q^2}{4\pi^2 \alpha_\text{em}} \left(\sigma^{\gamma^*}_T + \sigma^{\gamma^*}_L\right)$, and $ F_L = \frac{Q^2}{4\pi^2 \alpha_\text{em}}  \sigma^{\gamma^*}_L$. Here 
 the subscripts $T$ and $L$ refer to the transverse and longitudinal virtual photon polarization, respectively, and $\sigma^{\gamma^*}_{T,L}$ correspond to a sum of $q\bar q$ and $q\bar q g$ contributions. The experimental data is reported in terms of the reduced cross section
\begin{equation}
    \sigma_r(y,x,Q^2) = F_2\left(x,Q^2\right) - \frac{y^2}{1+(1-y)^2} F_L\left(x,Q^2\right),
\end{equation}
where $y=Q^2/(sx)$ is the inelasticity and $\sqrt{s}$ is the lepton-nucleon center-of-mass energy. 

\emph{Results} --- %
We calculate the proton reduced cross section $\sigma_r$ and the charm and bottom contributions to it ($\sigma_{r,c}$ and $\sigma_{r,b}$).
We use the NLO dipole-proton scattering amplitudes determined in~\cite{Beuf:2020dxl}, available at~\cite{heikki_mantysaari_2020_4229269}.
In particular, we use the ``light quark'' fits of~\cite{Beuf:2020dxl} where only the massless $u,d$ and $s$ quarks are included and the non-perturbative initial condition is fitted to the light quark contribution of the reduced cross section data measured at HERA~\cite{H1:2009pze}. This 
contribution is determined in~\cite{Beuf:2020dxl} by subtracting interpolated charm and bottom quark contributions from the total cross section data.  We do not include the fits to the inclusive HERA data 
as they use the massless quark cross sections to fit the inclusive data containing a substantial heavy quark contribution. 

In~\cite{Beuf:2020dxl} multiple different fits 
are reported, corresponding to different choices for the initial evolution rapidity $\Yobk$ and different schemes for the coordinate space running coupling and resummations of particular higher-order corrections. In total there are 12 fits reported for massless quarks. All different fits result in an approximately equally good description of the light quark contribution to the HERA structure function data.

\begin{figure}[tb]
    \centering
    \includegraphics[width=\columnwidth]{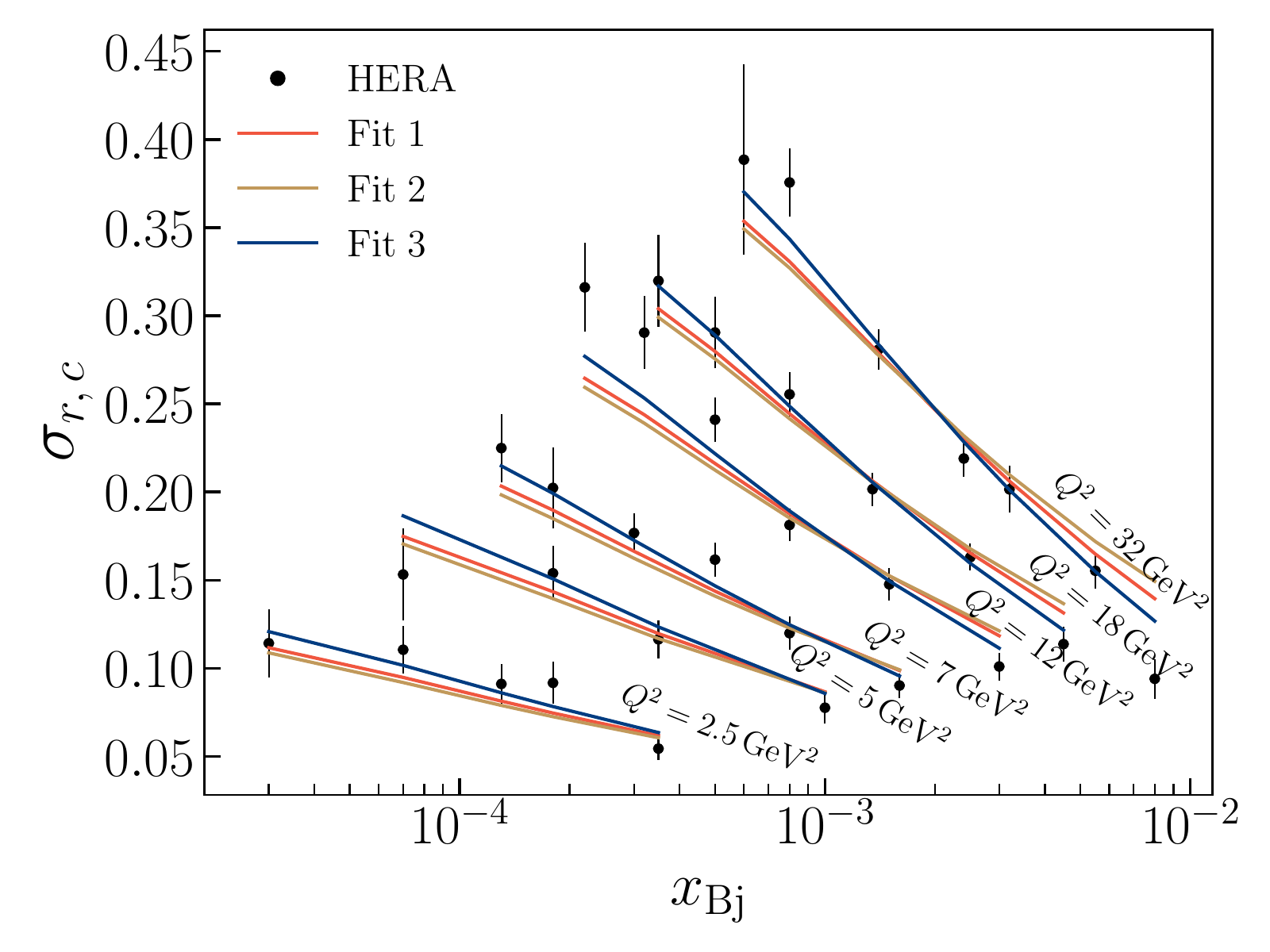}
    \caption{Charm reduced cross section predictions calculated using the different NLO fits from~\cite{Beuf:2020dxl} for the dipole amplitude that result in a good description of the charm data. The results are compared to the combined HERA data from~\cite{H1:2018flt}.}
    \label{fig:charm_sigmar_fit_uncert}
\end{figure}

We calculate predictions for the charm production cross section in the region $\xbj <0.01, 2.5\gev^2 \leq Q^2<50\gev^2$ using  all the different fits from~\cite{Beuf:2020dxl}, and compare the result to the HERA data from~\cite{H1:2018flt} in order to find which fits (if any) are allowed by the heavy quark production data. The charm mass (in the pole mass scheme used in the calculation of~\cite{Beuf:2021srj}) is allowed to vary within $1.1\gev<m_c<1.6\gev$. We consider a fit to be compatible with the HERA charm production data if 
one obtains $\chi_c^2/N \lesssim 2.5$ with the optimal charm mass. 
We find that predictions calculated by using three of the 12 fits are in excellent agreement with the charm production data.
This is illustrated in Fig.~\ref{fig:charm_sigmar_fit_uncert}, where a comparison to the HERA reduced cross section data in a few selected $Q^2$ bins is shown.
The H1 and ZEUS collaborations have also measured inclusive $b$ quark production
~\cite{H1:2018flt}, but due to the larger uncertainties and more limited kinematical coverage we do not use this dataset to determine which NLO fits from~\cite{Beuf:2020dxl} are allowed. We however note that using each of the three fits discussed above an excellent description of the $b$ quark production data is obtained. In each case we find $\chi_\mathrm{b}^2/N\lesssim 1.6$ when the $b$ quark mass is also fitted to this data.

The excellent agreement with the predicted heavy quark production cross sections
and the HERA measurements shows that at NLO it is possible to simultaneously describe all small-$\xbj$ proton structure function data. 
The results also demonstrate that the inclusion of the heavy quark production data to the extraction of the non-perturbative initial condition for the high-energy evolution equation provides additional tight constraints. 
Similar conclusions have also been made in calculations of exclusive heavy quarkonium production~\cite{Mantysaari:2022kdm,Mantysaari:2021ryb}. 
The advantage of the charm reduced cross section studied in this work is that one does not need to introduce an additional model uncertainty related to the non-perturbative vector meson structure.

The fits that are found to be compatible with the charm quark production data are summarized in Table~\ref{table:fits} along with the determined optimal heavy quark masses. 
The fact that the heavy quark data provides additional strong constraints for the determination of the initial condition for the BK evolution is expected. The heavy quark cross section is sensitive to much smaller dipoles than the inclusive one which can not discriminate fits that differ only at small dipole sizes. 
We note that the heavy quark production data only allows fits where the BK evolution is started at initial rapidity $\Yobk=0$. 
In the second class of fits considered in~\cite{Beuf:2020dxl} the dipole is frozen in the low-energy region $0<Y< \Yobk=\ln \frac{1}{0.01}$ where $Y$ is the evolution rapidity. This is not completely consistent as the $q\bar q g$ production cross section~\eqref{eq:sigma_qqg} in the soft gluon limit results in a (leading order) BK evolution for the dipole. 
Additionally, we note that (in the case of ResumBK and KCBK evolutions formulated in terms of the projectile rapidity) the parent dipole prescription for the running coupling is preferred.
We interpret that these physical constraints from the heavy-quark production data are because charm and bottom production probe dipole amplitudes in the perturbative region, and contribution from large dipoles dominating in light quark production with $N\sim 1$ is suppressed, see e.g.~\cite{Mantysaari:2018nng}.  
Based on the observations above we argue that the  fits summarized in Table~\ref{table:fits} are the ones that should be used in all NLO CGC calculations. 
The potential deviation between the predictions is then a measure of the model uncertainty after the non-perturbative input is constrained by all HERA structure function data.

\begin{table}[t]
\begin{tabular}{c||c|c|c|c|c|c|c|c||}
     \# & \begin{tabular}[c]{@{}c@{}}Resum.\\scheme\end{tabular} & $\as$ & $\Yobk$ & \begin{tabular}[c]{@{}c@{}}$m_c$\\$[\mathrm{GeV}]$\end{tabular} & $\chi^2_\mathrm{c}/N$ &  \begin{tabular}[c]{@{}c@{}}$m_b$\\$[\mathrm{GeV}]$\end{tabular} & $\chi^2_\mathrm{b}/N$ & $\chi^2_\mathrm{tot}/N$   \\
     \hline 
     1 & ResumBK & PD &  0 & 1.42 & 1.86 & $4.83$ & $1.37$ & 1.25\\
     2 & KCBK   & PD & 0 & 1.49 & 2.55 & $4.96$ & $1.58$ & 1.23 \\
     3 & TBK & BSD & 0 & 1.29 & 1.02 & $5.04$ & $1.12$ & 1.83 
\end{tabular}
\caption{Fitted initial conditions for the small-$\xbj$ evolution at NLO 
from~\cite{Beuf:2020dxl} that are compatible with the 
heavy quark production data from HERA. 
The corresponding 
charm and bottom 
masses are also shown. The terminology used to specify the resummation scheme and the running coupling prescription 
follows that of~\cite{Beuf:2020dxl}, and the abbreviation \emph{PD} refers to parent dipole and \emph{BSD} to Balitsky + smallest dipole~\cite{Balitsky:2006wa} running coupling. 
}
\label{table:fits}
\end{table}

\begin{figure}[tb]
    \centering
    \includegraphics[width=\columnwidth]{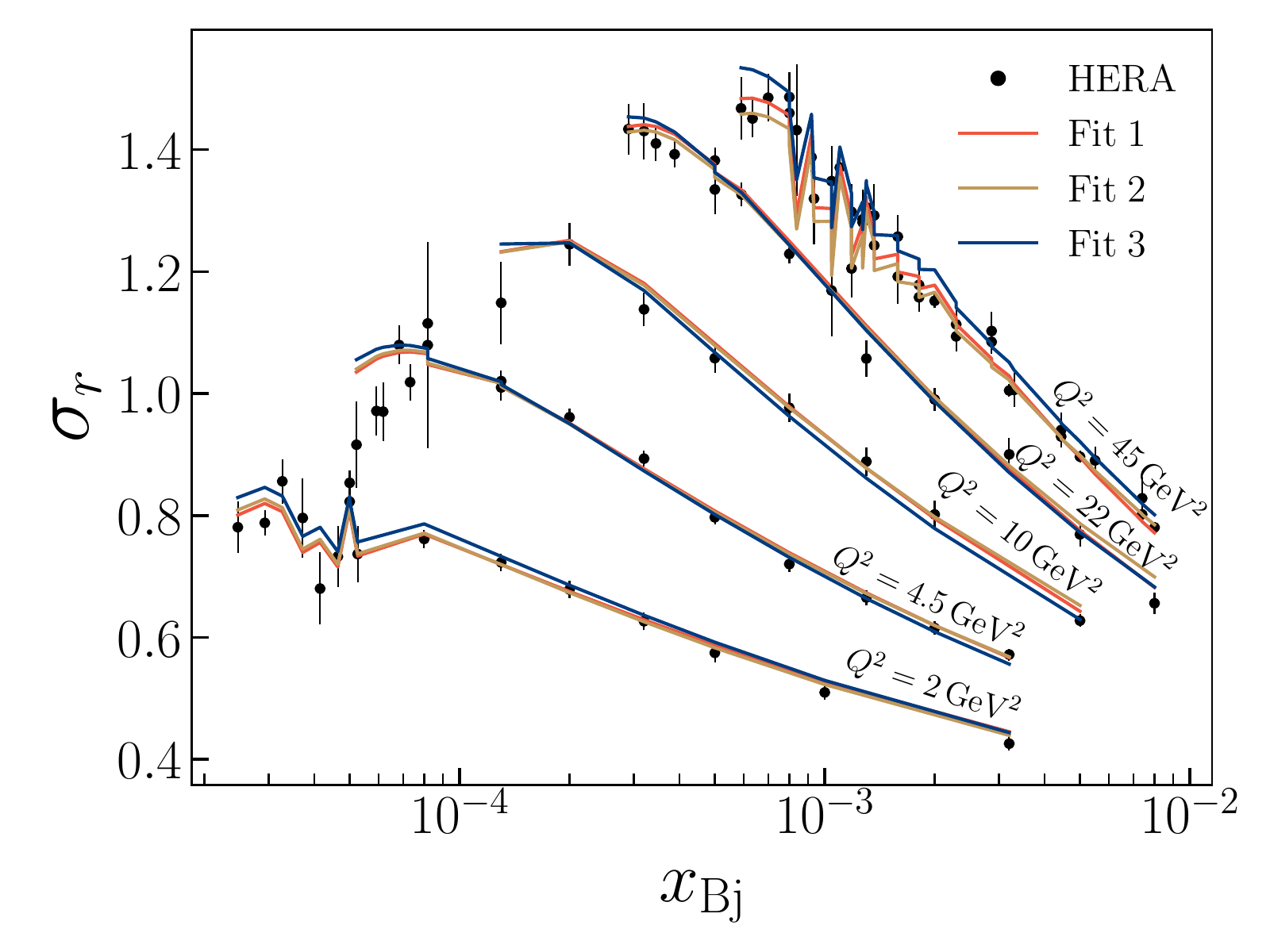}
    \caption{Total reduced cross section  
    calculated using the dipole amplitude fits allowed by the heavy quark production data.
    Note that as the $\sigma_r$ depends on inelasticity $y$, the 
    theory curves connecting the calculated points are not smooth.}
    \label{fig:total_sigmar}
\end{figure}

To more clearly illustrate the compatibility of the NLO CGC calculation with the most recent precise HERA data from~\cite{H1:2015ubc}, we show in Fig.~\ref{fig:total_sigmar} the total reduced cross sections computed using the dipole amplitude fits 
allowed by the charm 
data. 
We emphasize that this is the first time in the CGC framework that a simultaneous description of both the total and heavy quark production cross section is obtained when a perturbative small-$\xbj$ evolution equation is used to describe the center-of-mass energy dependence. 

Previous LO analyses have found it impossible to perform such a global fit to the HERA data without introducing, for example, additional parameters that render the proton probed by a charm quark dipole different from the proton probed by a light quark dipole~\cite{Albacete:2010sy}. 
A similar approximative NLO evolution equation as in this work was used in~\cite{Ducloue:2019jmy} but coupled to the LO impact factor. 
In that case, it was also found impossible to simultaneously describe the inclusive and heavy quark production data.

When the computation is  promoted to full NLO accuracy the mass dependence is modified for two reasons.
 First, after including higher-order corrections to the BK 
 equation (in projectile rapidity), the dipole amplitude does not anymore evolve towards an asymptotic shape with an anomalous dimension $\gamma<1$ (at small dipole sizes $r$ the amplitude behaves as $N\sim r^{2\gamma}$)~\cite{Albacete:2007yr}. Instead, the anomalous dimension (which is $\gamma \gtrsim 1$ in the fits reported in~\cite{Beuf:2020dxl}) remains approximatively constant suppressing the dipole amplitudes at small dipoles~\cite{Beuf:2020dxl,Lappi:2016fmu}. Hence,  the heavy quark production cross section is suppressed relative to light quark production. Second, adding the NLO corrections to the massive impact factor enhances the heavy quark production.
With TBK evolution we have opposite systematics: a small $\gamma$ is developed and the impact factor suppresses heavy quark production. 
 The net effect of these two competing NLO corrections
 is such that the mass dependence of the cross section matches that of the HERA data when the three fits identified in this work 
 are used.

\begin{figure}[tb]
    \centering
    \includegraphics[width=\columnwidth]{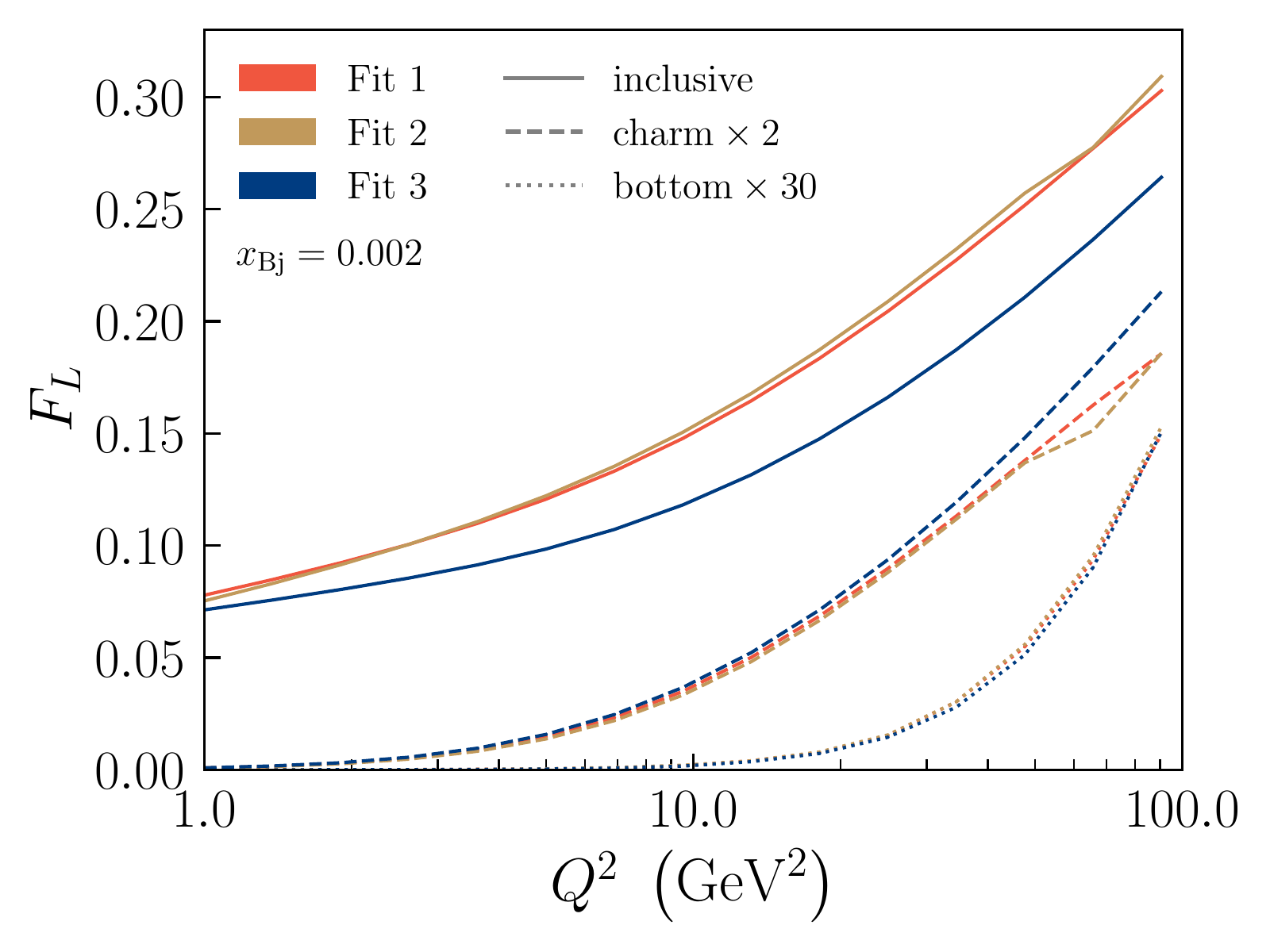}
    \caption{Total (solid lines), charm (dashed lines), and bottom (dotted lines) longitudinal structure functions as a function of photon virtuality in the EIC kinematics calculated using the three dipole amplitude fits compatible with the heavy quark data. 
    }
    \label{fig:fl}
\end{figure}

Finally, we illustrate the remaining theory uncertainty when performing NLO CGC calculations.
We  calculate predictions for the the proton longitudinal structure function $F_L$, and for the charm and bottom quark contributions to it, in the EIC kinematics. 
We take $\xbj=2\cdot 10^{-3}$, and show in Fig.~\ref{fig:fl} the  structure functions as a function of $Q^2$ calculated using the three 
fits determined above. 
For the bottom structure function, the 
different fits result in almost identical predictions for the EIC, whereas for charm production the predictions begin to differ  at $Q^2 \gtrsim 20 \, \mathrm{GeV}^2$. On the other hand, in the total longitudinal cross section a significant difference up to  $20\%$ is seen at all $Q^2$. Therefore, an inclusion of the future $F_L$ data in the global analysis will provide further constraints for the initial condition of the small-$\xbj$ evolution. The currently available $F_L$ data from HERA~\cite{H1:2013ktq} is not able to distinguish between the different fits.

\emph{Discussion} --- %
We have calculated heavy quark production cross sections in DIS at NLO in the CGC framework. The Bjorken-$\xbj$ dependence is obtained by solving the 
BK evolution equation with an initial condition extracted by fitting the total DIS cross section data in~\cite{Beuf:2020dxl}. We identify a small subset of the fits reported in~\cite{Beuf:2020dxl} that result in predictions for the  charm and bottom structure functions which are in excellent agreement with the HERA data~\cite{H1:2018flt}. These three fits, constrained by both the total and heavy quark cross section data summarized in Table~\ref{table:fits}, should be used in all future phenomenological applications at NLO accuracy. 

This is the first time in the CGC framework with 
perturbative energy evolution 
when a simultaneous description of all small-$\xbj$ proton structure function data is obtained. 
A good agreement with the HERA measurements is a crucial test for the gluon saturation physics incorporated in the CGC framework, and enables rigorous studies of non-linear QCD dynamics in DIS and other scattering processes. In particular, we demonstrate that 
global analyses including all small-$\xbj$ structure function data are feasible at NLO and that the heavy quark production data can provide additional constraints in such analyses.

As an application, 
we have calculated predictions for the proton longitudinal structure function $F_L$,  which will be measured accurately at the future Electron-Ion Collider. 
We reported predictions separately for the inclusive and heavy quark production cross section, and showed that the remaining model uncertainty is moderate. 
Including the $F_L$ data to the global analysis will further constrain the non-perturbative initial condition for the small-$\xbj$ evolution equations.

To fully explore the model uncertainties 
one should perform a global analysis to the HERA inclusive and heavy quark production data, taking into account the correlated experimental uncertainties, and extract the non-perturbative model parameters with their uncertainties directly from such an analysis. 
Additional constraints and more detailed probes of non-linear dynamics can be obtained by including other observables such as diffractive structure functions and exclusive cross sections.  
Such studies are becoming feasible thanks to the extensive progress toward NLO accuracy in the CGC framework, see e.g. Refs.~\cite{Boussarie:2016bkq,Caucal:2021ent,Mantysaari:2021ryb,Mantysaari:2022bsp,Mantysaari:2022kdm,Beuf:2022kyp,Bergabo:2022tcu,Bergabo:2022zhe,Caucal:2022ulg,Iancu:2020mos,Roy:2019hwr,Taels:2022tza}. In the future, we plan to perform a full Bayesian analysis to determine the likelihood distribution for all the model parameters, which will enable one to also fully take into account the propagation of model uncertainties.

\emph{Acknowledgments} --- %
We thank T. Lappi for discussions, and V. Apaja for computational support.
This work was supported by the Academy of Finland, the Centre of Excellence in Quark Matter and projects 338263, 346567, 321840, 347499 and 353772. This work was also supported under the European Union’s Horizon 2020 research and innovation programme by the European Research Council (ERC, grant agreement No. ERC-2018-ADG-835105 YoctoLHC) and by the STRONG-2020 project (grant agreement No. 824093).
J.P. is supported by the Finnish Cultural Foundation.
The content of this article does not reflect the official opinion of the European Union and responsibility for the information and views expressed therein lies entirely with the authors. Computing resources from CSC – IT Center for Science in Espoo, Finland and from the Finnish Grid and Cloud Infrastructure (persistent identifier \texttt{urn:nbn:fi:research-infras-2016072533}) were used in this work.

\bibliographystyle{JHEP-2modlong.bst}
\bibliography{refs}

\end{document}